\documentstyle[11pt,aas2pp4]{article}

\begin{document}
\newcommand{\etal}{\textit{et~al}}
\def\bibcode#1{}
\def\url#1{\texttt{#1}}
\def\astrobj#1{#1}
\def\PACS#1{PACS codes: #1}
\def\citeasnoun#1{\cite{#1}}

\title{Evidence for a Galactic $\gamma$-ray halo}

\author{D. D. Dixon}
\affil{University of California\\
        Institute for Geophysics and Planetary Physics\\
        Riverside, CA 92521}

\author{D. H. Hartmann}
\affil{Clemson University\\
        Department of Physics and Astronomy\\
        Clemson, SC 29634}

\author{E. D. Kolaczyk}
\affil{Department of Mathematics and Statistics\\
	Boston University\\
	111 Cummington Street\\
	Boston, MA  02215}

\author{J. Samimi}
\affil{Sharif University\\
        Teheran, Iran}

\author{R. Diehl, G. Kanbach, H. Mayer-Hasselwander, A. W. Strong}
\affil{MPE, 85740 Garching, FRG}

\begin{abstract}
We present quantitative statistical evidence for a $\gamma$-ray emission 
halo surrounding
the Galaxy.  Maps of the emission are derived.
EGRET data were analyzed in a wavelet-based non-parametric
hypothesis testing framework, using a model of expected diffuse (Galactic $+$
isotropic) emission as a null hypothesis.  The results show a 
statistically significant large scale halo surrounding the 
center of the Milky Way
as seen from Earth.  The halo flux at high latitudes
is somewhat smaller than the isotropic
$\gamma$-ray flux at the same energy, though of the same order
($O(10^{-7}$--$10^{-6})$ ph cm$^{-2}$ s$^{-1}$ sr$^{-1}$ above 1 GeV).  
\end{abstract}


\section{Introduction}
The study of diffuse high-energy $\gamma$-ray emission has proceeded
along two complementary paths.  The first is via the use of parametric
methods (see e.g. \citeasnoun{hunter}), where a physically motivated
model described by a small number of parameters is fit to the data.
The alternate approach (see e.g. \citeasnoun{chen} and \citeasnoun{willis})
is non-parametric, where one attempts to find the flux distribution
as a function of position without reference to a particular model.
The two classes complement each other in the sense that the strengths
of one are often the weaknesses of the other.  For instance, while
a parametric fit usually gives some global measure of how good the
fit is, but does not tell you where the model fails (e.g., ``the
model did not account for this blob of flux over there'').  The
non-parametric approach can give this information, but unlike the
parametric analysis, the quantitative assessment of the results
is complicated by the effects of statistical bias.  This latter
point should be born in mind when examining the results presented
in this paper.

Non-parametric analysis of photon-limited data is beset by
certain difficulties, key of which is that Poisson noise is
neither stationary nor additive.  Thus, the results of what would
be a straightforward analysis (e.g. smoothing by a Gaussian kernel) 
for data contaminated by white noise are more difficult to interpret.
We therefore apply a new wavelet-based technique which has the following
characteristics:
\begin{itemize}
\item{Rigorous treatment of Poisson statistics.}
\item{Assessment (in some sense) of the statistical significance
of the results.}
\item{Spatial adaptivity, in that structures at different size
scales are recovered automatically.}
\end{itemize}

In this paper we apply non-parametric analysis to EGRET data
taken during Phases 1-4 of the Compton
Gamma Ray Observatory (CGRO) mission.
Below we describe the non-parametric analysis method,
and present results from EGRET data showing an extended halo of $\gamma$-ray
emission apparently surrounding the center of the 
Milky Way.  We show that this halo
is statistically very strong in the data, and
not obviously attributable to any systematic effect of the analysis
or instrument, from which we conclude that it is most likely of astrophysical
origin.  We conclude with some brief discussion on the possible origins
and implications of the $\gamma$-ray halo.  This work expands upon
that first presented in \citeasnoun{dixon1} and \citeasnoun{dixon2}.

\section{Basics of the Method}
The non-parametric framework is a variant of the TIPSH (Translationally
Invariant Poisson Smoothing using Haar wavelets) methodology, first
introduced for denoising $\gamma$-ray burst light curves 
(\citeasnoun{kola1},\citeasnoun{kola2}).  
A detailed description of the approach is
described in a forthcoming paper (\citeasnoun{kola3}); we give a brief
outline here.

The goal of TIPSH is as follows: given a dataset consisting of observed
counts per pixel, estimate the {\em expected} counts per pixel over
the image.  In essence, we accomplish this by assuming that the 
photon intensity
distribution which generates the data contains local spatial correlations,
and that there are multiple length scales associated with these correlations.
That is, the pixel-to-pixel variations in the underlying photon intensity
distribution are not random, but rather have some structure 
related to the physics governing the emission.
{\em Wavelets} form bases for the space of square-integrable functions,
and are explicitly constructed to compactly encode such correlated
multiscale information.  We therefore expect that wavelets will be useful
for extracting the coherent information in favor of the random noise.
This expectation is borne out by a large body of work (see, e.g.,
\citeasnoun{donoho}).
In addition, wavelet transforms can be accomplished via fast algorithms,
which allow them to be realistically applied to larger datasets.

As implied by the name, TIPSH utilizes Haar wavelets.  These are the simplest
of all wavelets, simply taking the value $+1$ or $-1$ over the region
of non-zero support; 2D examples are shown in Figure~\ref{fig:haar}.
The Haar transform of an image provides a set of coefficients, one of
which is simply the average intensity, with the others providing information
about the intensity variation over different size scales in the horizontal,
vertical, and diagonal directions.  The particular form of the Haar wavelet
makes it conducive to analysis of Poisson data.  The sum of counts over
pixels is simply a Poisson variable, and thus (excepting the DC
level) the Haar coefficients are distributed as the {\em difference} between
two Poisson distributed random variables.  This, coupled with the
statistical independence of the coefficients within a given size scale
and direction
(the wavelets have no overlap), allow us to derive convenient expressions
for the statistical distribution of wavelet coefficients
(see \citeasnoun{kola3} for more
details).  We can then use these to calculate the probability that
a coefficient of the observed size or larger is statistically
consistent with some null hypothesis.

TIPSH is a ``keep-or-kill'' strategy, applied to the Haar coefficients
of the difference between the data and some hypothesized count distribution,
i.e., an estimate of the background or ``known'' emission.  
If the magnitude of the coefficient exceeds some threshhold,
it is kept, otherwise it is zeroed out.  After application of this procedure,
the inverse Haar transform is applied to the remaining coefficients,
giving the estimated denoised residual count distribution.  Except for the
case of a constant background, simple expressions for the keep-or-kill
threshholds are not available.  We instead exploit the following property:
given threshhold $t$, and an observed value $h^{\mathrm obs}$ of wavelet
coefficient $h$,
\begin{equation}
h^{\mathrm obs} > t \quad\hbox{if and only if}\quad 
\Pr(h > h^{\mathrm obs}) < \Pr(h > t)\enskip .
\label{eq:pval}
\end{equation}
Calculation of these probabilities (known as {\em p-values}) requires
the specification of a Poisson distribution under some null hypothesis
or ``background''.  We supply a count distribution which serves
as the null hypothesis.  We then need a way of specifying the p-value
cutoff corresponding to a ``significant'' detection.  For this paper,
we use the level-wise false detection rate (FDR), which is simply the
probability that one or more wavelet coefficients in a given
resolution and direction would accidentally be deemed significant
if the null hypothesis reflected the actual intensity distribution
underlying the data.
The wavelet coefficients within a given size scale and direction
(horizontal, vertical, diagonal) are statistically independent, 
as they have no spatial overlap.  
The coefficient-wise FDR $\alpha_k$ for the $k$th coefficient 
is thus simply calculated from the expression
\begin{equation}
{\mathrm FDR} = 1 - (1 - \alpha_k)^n,
\end{equation}\label{eq:coeff}
where $n$ is the number of coefficients for a particular size scale.
We calculate $\Pr(h_k > h_j^{obs})$ via an algorithm due to
Posten (\citeasnoun{posten}), use the relation $\Pr( h_k > t) = \alpha_k/2$,
and apply eqn.~\ref{eq:pval} to decide if we keep coefficient $k$.

Finally, we note that the analysis is carried out in the translationally
invariant or ``cycle spinning'' (\citeasnoun{donocoif}) framework.  The main
reason for this is that the Haar wavelets themselves are not smooth
functions, while we expect a somewhat higher degree of regularity
in the estimated intensity distribution.  More specifically, the Haar
function is piecewise constant and (in the continuum limit) non-differentiable.
Astrophysical intensity distributions, however, are likely ``smoother''
in the sense of being locally approximated by higher order polynomials.
The compression property of the wavelet transform is directly related
to this.  Wavelets are often explicitly constructed to be orthogonal
to polynomials up to some given order.  As such, regions in a signal
which are closely approximated by polynomials of that order or smaller
will yield small wavelet coefficients.  The Haar wavelet is the lowest
order wavelet in this sense, being orthogonal only to the constant
function.  Though the (discrete) Haar basis can be used to represent
{\em any} (discrete) measured image or intensity distribution,
some information is spent to ``smooth out the corners'', i.e., to
fill in any higher order smoothness.  The threshholding
procedure is likely to wipe out some of that information, resulting
in pseudo-Gibbs artifacts reflecting the ``blocky'' appearance of
the Haar wavelets.  While going to a higher order wavelet nominally
solves this problem, in this case adjacent wavelets would have spatial
overlap, and so their coefficients would no longer be statistically
independent.  Further, the wavelet coefficients in this case are
constructed from non-trivial weightings of measured counts, and as
such the corresponding p-values are much more difficult to calculate.
So the use of Haar wavelets is necessary to maintain the rigor and
simplicity of our approach.

To mitigate the functional/approximation shortcomings of Haar wavelets,
we instead effectively perform the analysis over all horizontal
and vertical shifts of the data, and average together these results
to obtain the final estimate.  In actuality, much of the information
in such a procedure is redundant, and so we can obtain the requisite
information via an $O(N \log N)$ algorithm.  Without going into
the mathematical details (see \citeasnoun{donocoif} for a discussion),
the average-over-shifts procedure yields gains in both the continuity
and approximation order of the estimate when compared with the
classical Haar basis.  An analogous astronomical procedure is that
of {\em dithering}, where multiple observations of the same region
are taken while moving the telescope by an amount smaller than
the angular size of a CCD (or other) pixel.  We emphasize that
statements about detection rates, etc., apply only {\em within}
a given shift.

\section{Details of the Analysis}
The data analyzed and discussed below
are for $E>1$ GeV, from Phases 1--4 of the CGRO
mission.  For completeness, we shall later present results for the
energy ranges $100<E<300$ MeV and $300<E<1000$ MeV, but our discussion
here will concentrate above 1 GeV.
Analysis is performed directly on the photon data, binned
in $0.5^\circ\times 0.5^\circ$ pixels.
The wavelet transform is most easily implemented for datasets
of size $2^J$, where $J$ is some integer.  The nominal all-sky
EGRET dataset is $720\times 360$ pixels.  To achieve the proper
size, we pad the $720\times 360$ data to a size of $1024\times 1024$
using segments of the data to achieve spherical boundary conditions.
This implies that the data are periodic in longitude, while the pixels
at the poles are aligned with their counterparts 180 degrees away.
In this way the proper symmetries of the celestial sphere are enforced,
reducing the possibility for artifacts due to artificial
discontinuities at the data boundaries.  Similar padding was performed
on the null hypothesis, described below.

The null hypothesis consists of the sum of two components.  The first
is uniform in flux, accounting for the isotropic $\gamma$-ray background,
and is included at a level computed from
the spectrum of \citeasnoun{sreekumar}.
The second component is standard Galactic diffuse model used in
EGRET likelihood analysis, which is publicly available from the
Compton Observatory Science Support Center (COSSC);
this model is shown in Figure~\ref{fig:diffmod}.
This model is similar to that described by \citeasnoun{bertsch}
and \citeasnoun{hunter}, and describes $\gamma$-ray emission due
to cosmic-ray interactions with interstellar matter and low-energy
photons via the processes of bremsstrahlung, inverse Compton scattering,
and nuclear processes ($pp\rightarrow \pi^{\pm,0},
\pi^0 \rightarrow \gamma\gamma$).  The predicted $\gamma$-ray
intensity is computed from line-of-sight integrals for these
processes, based on an estimate of the three-dimensional distribution
of Galactic matter, some cosmic-ray injection spectra, and the assumption
that the cosmic-ray source density is (at least on large scales) closely
correlated with the distribution of matter.  The diffuse model obtained
from COSSC has been convolved with the approximate EGRET PSF for
a given energy band.  The predicted count distribution is
given by the sum of the Galactic and isotropic 
components multiplied by the composite
exposure factor of Phases 1--4.

Each $1024 \times 1024$ padded dataset was processed by TIPSH, using the null
hypothesis described above and a variety of FDR's.  Maps were
generated for FDR's ranging from $10^{-2}$ to $10^{-15}$, and though
the map details changed (the map becomes smoother as the FDR is decreased), 
the basic result we present in this paper
was essentially unchanged.  Remember that the FDR we use applies to
an entire resolution level and particular shift and direction, and gives the
probability that if the null hypothesis were true, we would
keep one or more wavelet coefficients due solely to Poisson fluctuations
in the data.  Reference to eqn.~\ref{eq:coeff} shows then that
the coefficient-wise FDRs $\alpha_k$ are going to be considerably
smaller than the level-wise FDR.  
The particular level-wise FDR used for the figures is 
$10^{-4}$.  One can envision other choices for the level-wise
(or coefficient-wise) FDRs; the choice of a single level-wise FDR
is easy to implement and makes for straightforward interpretation.

\section{Results and Consistency Checks}
The TIPSH residual relative to the standard diffuse prediction
is shown in Figure~\ref{fig:halo}.
This shows the all-sky map for $E>1$ GeV generated by the processing
described above.  For clarity, we have separated the positive and
negative components of the residual.  
The positive residual map shows several localized excesses corresponding
to known point sources, as well as a general excess in the Galactic
plane which is discussed in \citeasnoun{hunter}.  Further, we
can clearly see what appears to be a large scale emission halo 
surrounding the Galaxy.  The negative residual
map shows only a few artifacts related to point sources; these are
expected since wavelets, while providing an excellent representation
of extended spatially-correlated structure, perform rather poorly for
isolated localized structures such as point sources.

The presence of point sources in the residual
is expected, since we did not account for the point sources
in our hypothesis.  The plane excess is also expected, as it is known that
the spectral distribution of the standard Galactic diffuse model
underpredicts what is observed above 1 GeV. \citeasnoun{hunter}
states (for a slightly different version of this model) that the
Galactic plane emission is actually 60\% greater than that predicted.
\citeasnoun{pohl} gives a figure of a 40\% excess over the whole
sky, and further claims that a simple scaling of the standard
model will account for the excess.

To better understand which parts of the residual can be accounted
for by such scaling and which represent a truly different spatial
distribution of flux, we reanalyze the data using hypotheses where the
Galactic diffuse contribution has been scaled up by some factor.
The result for a 40\% increase is shown in Figure~\ref{fig:40}.
From the largely negative regions along the plane, we see that the
contribution from the standard model distribution has been overestimated
in this case.  Yet high latitude excess is still apparent about GC,
indicating another component whose spatial distribution is {\em not} 
correlated
with the model.  Figure~\ref{fig:20} shows the result for a scaling
of 20\%.  Here we see little systematic excess or deficit along the plane,
with much of the high latitude emission away from GC also having been
removed.  However, a large-scale feature centered at GC is still
clearly evident.  Our claim is that this feature cannot be made
consistent with the spatial distribution predicted by the standard
diffuse model.  Since the 20\% scaling best isolates the anomalous
halo component, we shall refer to this result in what follows.

The flux levels shown in Figure~\ref{fig:20} are in the range
$10^{-7}$--$10^{-6}$ ph cm$^{-2}$ s$^{-1}$ sr$^{-1}$.  We shall
further discuss below the various potential systematic errors
in this estimate, and here simply note that at least as regards
the large-scale structure, the flux estimate is probably good
to a factor of two, if one neglects regions of low exposure.
Unfortunately, estimates of statistical errors are generally
difficult to obtain for non-parametric analyses, and this is the
case here as well.  We can avoid some of the difficulties related to
exposure systematics by directly examining the counts.  Of the
$\sim$85000 counts in the dataset, $\sim$71000 are accounted
for by the null hypothesis, including the 20\% scaling of the
Galactic diffuse model.  From the GeV source catalog of \citeasnoun{lamb}
(which uses data similar to what we have employed), we estimate the
total point source contribution to be about 7100 photons, leaving
about 7400 in the rest of the residual, which appears to be
primarily associated with the halo.

The immediate question arises as to the statistical significance of
this feature.  Though we are able to make rigorous statements about
the coefficient-wise and level-wise FDR, similar quantification of
object-wise significance (e.g., ``this blob is significant at the
$n\sigma$ level'') are difficult.  The main reason for this is that
a given object, such as the halo, is most likely composed of many
wavelets which are not statistically independent, and at this point,
estimates of object-wise error rates appear to be mathematically
intractable.  However, while we can't assign a number to it, we can argue
that the feature is quite significant.  First, we may examine how
many wavelets were found to be inconsistent with the null hypothesis,
and compare this with our expectation for statistical fluctuations.  
Results
are shown in Table~\ref{tab:rates}.  Our expectation is that if the
null hypothesis were accurate, only 0.01\% of 
the time would we accidentally detect a wavelet as significant due
to noise.  A large scale structure such
as the halo is clearly composed mostly of large wavelets, and we find
that the actual detection rates
for the large scale wavelets greatly exceed the
expected rate for noise.  Second, the halo persists even at a FDR of
$10^{-15}$, while most other features (such as point sources) are
removed at such a severe threshhold.  Remember that this measure of
``significance'' applies to the {\em wavelet} representation of the
residual, and so we expect structures with a large degree of
spatial correlation to survive harsher threshholding better
than highly localized features (e.g., point sources).  The opposite
would be true if we were to perform threshholding directly on the pixel
count values.  Thus the halo clearly is a very strong 
effect in the data in terms of being a spatially coherent (albeit
dim) structure.

The assessment of statistical significances is one of those areas which
is difficult in non-parametric analyses, but quite straightforward in
a parametric analysis.  So we may also address the significance of the
halo by fitting some model and examining the likelihood ratio when
compared to a model that does not include a halo component.  Clearly the
exact result here is model dependent, but this approach at least gives
us some handle on the statistics.  For the case at hand, we fit a simple
linear model:
\begin{equation}
\bar{d}_{ij} = E_{ij}(\alpha\cdot \mathrm{ISO}_{ij} + \beta\cdot 
	\mathrm{GDM}_{ij} + \gamma\cdot \mathrm{IC}_{ij}),
\end{equation}
where the subscripts refer to the $(i,j)$th pixel, $E$ is the exposure
factor, ISO is the predicted isotropic background, GDM is the predicted
Galactic diffuse model, and IC is a particular model of inverse Compton
emission derived from a cosmic-ray propagation code (\citeasnoun{strong4}).
The distribution of IC is shown in Figure~\ref{fig:ic}a, and while
it may or may not reflect the ``true'' distribution of the halo, it at
least gives us some component with approximately the right spatial
characteristics.  The parameters $\alpha,\beta,$ and $\gamma$ are
estimated by maximizing the Poisson likelihood.  The null hypothesis for
our likelihood ratio test simply omits the IC component.  The full fit
gives values $\alpha = 0.975$, $\beta = 1.3286$, and $\gamma = 0.5311$,
while the fit to the null hypothesis gives $\alpha = 1.2428$ and
$\beta = 1.4352$.  Following \citeasnoun{mattox}, we find a formal
signficance of
\begin{equation}
\sqrt{TS} = \sqrt{2(\log L - \log L_0)} \simeq 21\sigma.
\end{equation}
Remember that this is the level at which the null hypothesis is
{\em rejected} given the specific test we performed, and says nothing
about how well the specific morphology of IC compares
with some other morphology.

Processing by TIPSH gives an estimate of the average residual counts/pixel
We then divide by the exposure to give the estimated flux over the
region.  This point merits some discussion.  The total EGRET exposure
is highly non-uniform on the sky, tending to be concentrated on sources
of interest (see Figure~\ref{fig:exposure}).  
Thus, a spatially uniform flux signal yields a highly 
non-uniform
count distribution in the data.  TIPSH basically performs adaptive smoothing
in count space, and though the algorithm is count conserving (w.r.t. the
{\em total} counts in the dataset), details
of the exposure pattern can get smoothed over, especially in areas of
low exposure.  Therefore we need to be careful in our interpretation of
flux maps, in that we expect some anticorrelation
of map features with the exposure pattern.  These systematics generally
also lead to distortion in the flux levels, so the contour levels
shown in Figures~\ref{fig:halo}, \ref{fig:20}, and \ref{fig:40}
should be taken with a grain of salt.
Again, this effect is most pronounced in low exposure and/or low count 
regions.  For comparison, Figure~\ref{fig:haloexp} shows the exposure
overplotted on the residual flux map of Figure \ref{fig:20}.
Similarly, while TIPSH is count-preserving, the distortions due to the
spatially non-uniform exposure imply that the total map flux is systematically
off.  We find that the total residual flux is usually overestimated,
though generally only by a factor of two or so.

To better illustrate some of the statistical and systematic effects
described above, we present some examples from simulated data.
We begin showing the results of running TIPSH against a dataset
generated directly from the null hypothesis (Galactic diffuse
$+$ isotropic).  Our expectation is
that the level-wise FDR of $10^{-4}$ should effectively suppress
the noise, with very few wavelets passing threshhold.  This
expectation is borne out by the actual detection rates in
Table~\ref{tab:rates}, as well as the residual shown in
Figure~\ref{fig:sim0}.

Our second set of examples illustrates the level of systematic
errors induced by non-uniform exposure effects.  As we shall see,
the spatially non-uniform exposure is the limiting factor in
assessing morphology and flux levels.  A simple way to see this
effect is to add in excess isotropic signal, and examine the
structure returned in the flux map.  We thus simulate data
with an isotropic component twice that which is measured, and
use the same null hypothesis as above.  Figure~\ref{fig:flat}
shows the fractional deviation 
(given as (residual - isotropic)/isotropic)
from the expected isotropic signal
in the resultant residual flux map.  As expected, we see anticorrelation
with the exposure pattern (overplotted and also shown in
Figure~\ref{fig:exposure}).  Next we simulate data for with
a 40\% excess in the Galactic diffuse contribution.  The input
excess and TIPSH residual are shown in Figure~\ref{fig:sim40}.
The level of distortion here is not so drastic, but again this
might be expected since most of the excess is concentrated near
the Galactic plane where the exposure tends to be relatively
large.  Figure~\ref{fig:sim40}b lends further credence to our
argument that the halo is not simply an artifact of the standard
model's underestimation of the observed flux above 1 GeV.  Finally
we simulate data where we include the IC model used in our
parametric fit above, to see the effect of a more ``halo-like''
excess.  Figure~\ref{fig:ic}b shows the TIPSH residual.  While
there is some loss of small-scale information and obvious artifacts
in regions of low exposure, we note that the contours corresponding
to the broader components of the emission are not overwhelming
displaced, particularly in the neighborhood of $l = 0^\circ$.
A similar conclusion might be drawn from Figure~\ref{fig:icrat},
which shows the ratio of the model to the recovered residual.  This
should not be too surprising, perhaps, since coefficients corresponding
to large scale wavelets are derived from larger numbers of photons, and
we thus expect them to be statistically more robust than their smaller
scale counterparts.

The above discussion has largely concentrated on the spatial distribution
of the halo above 1 GeV.  We close this section by briefly touching
on the spectral characteristics.  Though we leave detailed analysis
of the energy dependence to a future paper, we show the TIPSH results
for 100-300 MeV and 300-1000 MeV in Figures \ref{fig:100} and
\ref{fig:300} respectively.  These are derived using the unscaled
standard Galactic diffuse model and isotropic fluxes for these
energy ranges derived from \citeasnoun{sreekumar}.  Note that, unlike
the case above 1 GeV, if anything the plane flux in these cases
is {\em overestimated} by the 
model, and that the halo excess is still clearly present.  It is tempting
to simply read off fluxes from the various figures and construct a
spectrum, but the discussion above certainly indicates that such an
exercise should be undertaken with caution.  Detailed spectral analysis
necessarily requires a parametric approach, and we leave this for
future research.

\section{Discussion of potential systematics}
Let us address some possible systematic non-astrophysical sources of the halo:
\begin{itemize}
\item{{\em Exposure related}--From the discussion above, we do
expect some artifacts due to the non-uniform exposure when the
map is converted from counts to flux.  However, the exposure
variations occur on much smaller angular scales than that of the
halo, and of course there's no artifacts without some excess
flux in the first place.  The examples from simulated data
in the previous section certainly indicate that we are seeing
a non-isotropic halo-like excess, albeit distorted by exposure
artifacts.
Another exposure related possibility might be expected
from the threshholding behavior, in that more threshholding is
going to occur in low-exposure regions, where there are less
statistics.  One might naively expect larger signal suppression
in regions of lower exposure, as is sometimes seen, for example,
with Maximum Entropy, 
resulting in apparent excess emission following
the exposure map.  Wavelets, however, have zero mean, so threshholding
a particular wavelet does not change the total counts in the region
encompassed by the wavelet support, but
rather removes local structure variations at that scale (i.e., TIPSH is
count-preserving).  So, signal is not lost, but rather smeared out over
a larger area in low-exposure areas, leading to the kind of smaller-scale
artifacts discussed above.  Examination of Figure~\ref{fig:haloexp}
shows no obvious large-scale correlation of the halo and exposure; for example,
the Galactic center and anticenter regions both have significant exposures,
but the excess appears mostly about the Galactic center.}

\item{{\em EGRET calibration errors}--It is known (\citeasnoun{sreekumar})
that the EGRET spark chamber efficiency varies over time, due to
degradation in the gas, gas refills, etc.  Calibration errors
in the time-variable efficiency could conceivably lead to the appearance
of apparently diffuse features.  However, to generate what we see in
Figure~\ref{fig:20} would either require a very large error
(on the order of 50\% or greater) in calibrating a single observation
near GC, or a systematic underestimation in the efficiency
of most observations near
GC, but nowhere else on the sky.  Both of these options seem unlikely.
Another systematic problem may come from
the exposure at large detector zenith angles, which is not well 
quantified.  The exposure at these large angles is
small, and should not contribute much in a composite all-sky map.  Further,
we have performed the same analysis on datasets with a $30^\circ$ zenith
cut, and again find that the large-scale results are essentially
unchanged (additional small-scale artifacts appear due to the ``hard edge''
in the exposure caused by the zenith cut).}

\item{{\em Errors in the point spread function}--Since we are
denoising and not deconvolving, the point spread function (PSF) only
enters into the analysis through the Galactic diffuse model, where the
predicted $\gamma$-ray flux profile is convolved with the PSF to
generate the expected intensity distribution in the data.  The PSF
used in this calculation is zenith averaged, so some small local
deviations from the true PSF are likely.  Such errors are
probably not significant except for the brightest of sources, and
certainly wouldn't generate the type of large-scale structure we see
here.  One could also posit a low-level spatially extended tail in
the PSF.  It is rather difficult to think of a physical cause for
this ($\gamma$ and $e^\pm$ scattering in the instrument are 
presumably well understood), and in this case 
the large-scale longitudinal smearing of Galactic plane emission would be
considerably greater than what we see, which is of a rather more
spherical appearance in Figure~\ref{fig:20}.}

\item{{\em Oversmoothing of Galactic plane flux}--Denoising invariably
requires some level of data smoothing, and so structures deemed
significant by the algorithm generally appear somewhat smeared out,
with the degree of smearing depending upon the statistical significance
(less significant $\rightarrow$ more smeared).  The locally adaptive
nature of TIPSH should mitigate this effect a great deal, when compared
with a non-adaptive method such as simply blurring the data with a
Gaussian.  Nonetheless, we should address the possibility that
some fraction of the excess Galactic emission above 1 GeV is being 
smoothed out to high latitudes, and giving the appearance of a halo.
We can easily argue against this possibility by examining Figure~\ref{fig:40}.
Here, we have a general {\em deficit} in the Galactic plane, but
still detect a halo excess.  Since all of the
smoothing occurs in the residual, i.e., none is applied to the hypothesis,
we would not see the halo without a plane excess if 
oversmoothing were the culprit.}

\item{{\em Particle background}--While EGRET does use a charged
particle anticoincidence shield, potential for proton-initiated background
contamination does exist (see \citeasnoun{sreekumar} 
for discussion).
Further detailed study using Monte Carlo and examination of individual
event track data is necessary to rule this out completely.  However,
we note that to date, there is no indication that cosmic-ray
protons show such strong spatial anisotropy as seen here
(see, e.g., \citeasnoun{longair}).  Further, the signal appears correlated
with the Galaxy, as opposed to any particular orbital characteristics,
and shows similar average properties as was seen
in the COS-B result (\citeasnoun{strong2}).  
A non-$\gamma$-ray neutral particle
signal seems even more unlikely.}

\item{{\em Contamination by Earth albedo}--Rejection of Earth albedo
is accomplished by making a cut on the reconstructed
$\gamma$-ray arrival direction, based on the position of the Earth within
the FOV.  Due to the finite instrumental PSF, this approach is not perfect,
and we expect some albedo contamination (\citeasnoun{willis}).
We emphasize that albedo contamination is
largely a {\em local} effect, with the key question being the location
of most of the contamination.  Figure 3.6 of \citeasnoun{willis} shows
a plot in celestial coordinates of those regions dominated by exposure
to large ($\alpha > 80^\circ$) Earth-limb zenith angles, for which it
is expected that albedo contamination will be most pronounced.  We note
that those regions with the largest expected albedo
contamination are confined to the celestial polar 
regions, with $|\mathrm{dec}|\geq 50^\circ$.  Curves delineating these
regions on a Galactic plot are shown in Figure~\ref{fig:albedo} overplotted
on the residual flux map.  While some parts of the halo do
appear within these regions, most of it lays outside, and there
is clearly no correlation between the halo distribution and the location
and/or size of these areas.  From this
we conclude that the halo is not due to albedo contamination.}
\end{itemize}

Ultimately, the only definitive statement we can make is that the
halo is ``in the data'', statistically speaking.  While $\gamma$-ray
data is rather prone to contamination by systematic effects,
given the above discussion and results, we
feel the most likely explanation of the halo is that it is astrophysical
in origin.

\section{Comparison with Previous Results}
The idea of a large-scale, yet non-isotropic component of Galactic 
$\gamma$-ray emission is not new.  
\citeasnoun{strong1}
provides evidence
for such a feature from the COS-B data after subtracting the estimated
contribution from cosmic-ray/gas interactions.  In 
\citeasnoun{strong1},
it is noted that the effect is larger in the inner Galaxy, suggesting
a Galactic, rather than local origin, with inverse Compton (IC) proposed
as the emission mechanism.  \citeasnoun{strong2}, 
however, investigates a
possible local origin as well, citing the apparent north-south
asymmetry of the emission as an indicator of a local source.
\citeasnoun{chen} also visit this problem by
calculating the linear correlation of the $\gamma$-ray emission with
the HI column density in eight large regions with
$29.5^\circ < |b| < 79.5^\circ$ and noting that the y-intercept of
the linear fits (``uncorrelated emission'') shows some anisotropy
in longitude.  These authors also propose an IC origin,
fitting a model which includes the spatial distribution of 408 MHz 
emission, on the assumption that this
traces the cosmic-ray electron distribution.  
\citeasnoun{smialkowski} perform
a similar analysis, and also show evidence for a high-latitude excess.
\citeasnoun{wright} claims a weak anisotropy in the diffuse
$\gamma$-ray background.
\citeasnoun{willis} shows evidence for a $\gamma$-ray
halo in EGRET data for $E>100$ MeV by examining the the residual above
the Maximum Likelihood model, consisting of known point source contributions,
expected emission from
cosmic-ray/matter interactions, a model for IC emission, and the isotropic
$\gamma$-ray background.  The residual, blurred by a Gaussian of width
$5^\circ$, shows evidence for large-scale non-isotropic emission
surrounding GC.  Willis also constructs a ``large-scale residual'' by
estimating the flux at every point on the sky from a fit only to data within
$15^\circ$ of that point, and then subtracting the predicted Galactic
diffuse emission.  \citeasnoun{sreekumar} 
also note evidence for
an extended excess about GC, following an analysis similar to that
of \citeasnoun{willis}.

If we compare our results with those of previous investigators,
the case for an astrophysical origin of the halo is further strengthened.
Previous authors (\citeasnoun{strong1}; 
\citeasnoun{strong2}; 
\citeasnoun{chen}; 
\citeasnoun{smialkowski}; \citeasnoun{willis}) 
have noted the following characteristics:
\begin{enumerate}
\item{Anisotropic distribution of $\gamma$-ray flux above that implied
by gas column densities.}
\item{A longitude distribution which appears to reach a maximum near
the Galactic center.}
\item{A north-south asymmetry in the latitude distribution, with more
flux appearing at positive latitudes.}
\end{enumerate}
These characteristics are also reflected in Figures \ref{fig:halo},
\ref{fig:40}, and \ref{fig:20}.  
That the different analyses of EGRET data give
similar results is maybe not such a strong statement, as they all
basically look at residuals not correlated with gas, and
would be subject to similar instrumental systematics.  However,
the similarity amongst EGRET results certainly argues that we are
not seeing artifacts of any individual analysis.  That
COS-B sees the same basic characteristics 
(\citeasnoun{strong1}) provides a more
compelling argument for an astrophysical origin.  Most notably, the
orbit of COS-B was quite different than that of CGRO, being at $90^\circ$
inclination (vs. $28.5^\circ$), with a much larger perigee and eccentricity.  
If the halo and its characteristics were due to any sort of orbital or
background systematics, it would seem a highly unlikely coincidence.

\section{Discussion of potential astrophysical sources}
If the halo is astrophysical, the next question is as to the source of
the $\gamma$-rays.  The possibilities are to some extent dependent on
what we take as the ``distance'' to the halo, i.e., is it local or
associated with the Galaxy on a large scale?  The particular location
and large-scale morphology certainly suggest the Galactic interpretation;
a local feature with such characteristics would be highly coincidental.
As suggested by \citeasnoun{strong2}, however, 
the latitude asymmetry may
argue for a local origin. It is also possible that we are observing
multiple phenomena along the same lines of sight.
In the absence of arguments for a local origin (e.g., nearby molecular
clouds with similar location and size), we shall focus our discussion on
possible Galactic origins.

As discussed above, determination of the precise extent of the halo is
hampered by the statistics and non-uniform exposure of the dataset, but
if we take it as Galactic, then it clearly extends several kpc above
and below the Galactic plane.
Three obvious possibilities present themselves for an extended halo 
$\gamma$-ray source distribution:
\begin{enumerate}
\item{Unresolved high latitude point sources}
\item{Inverse Compton emission}
\item{Gamma rays associated with particle dark matter (either baryonic or not)}
\end{enumerate}
This list is clearly not exhaustive, but represents possibilities which
we have examined.

A halo population of sufficiently dim and numerous point sources might
account for the observed $\gamma$-ray excess.  The obvious candidates here
would be \astrobj{Geminga}-type pulsars, presumably ejected from the 
Galactic plane
via asymmetric supernova explosions.  Such scenarios have been discussed
extensively in the context of $\gamma$-ray bursts.
Starbursts at the Galactic center may have injected a large number of
dimming pulsars into the Galactic halo 
(\citeasnoun{hartmann}).
One can argue in favor of such an explanation in terms of the characteristics
of \astrobj{Geminga}, e.g., nearly all of its luminosity is in the form of
$\gamma$-rays.
To constrain this scenario, one would need to generate an estimate
for the total inegtrated (over $4\pi$ sr) flux from our halo, and
assume $10^3$-$10^5$ \astrobj{Geminga}-like pulsars.

One might also postulate some other
class of high latitude point sources which may generate the halo
$\gamma$-ray emission.  An intriguing possibility is that of
primordial black holes (PBHs).  As they evaporate, PBHs would
produce $\gamma$-rays, leading to a diffuse $\gamma$-ray background
\citeasnoun{page}.  Clustering of PBHs around galaxies would lead to
anisoptries in this background, most notably as observed from the
Milky Way.  \citeasnoun{wright} provides evidence for a weak anisotropy,
and places limits on the PBH evaporation rate.  \citeasnoun{cline}
discusses this further, and notes that an extended halo population
of PBHs could lead to a halo such as we observe, with a significant
portion of the isotropic $\gamma$-ray background also due to PBH
evaporation in the extended halo.

The possibility of high latitude inverse Compton (IC) emission has been
discussed previously (see, e.g., 
\citeasnoun{smialkowski}, 
\citeasnoun{chen},
\citeasnoun{strong2}, \citeasnoun{stryou}, 
\citeasnoun{willis}).
The distribution of Galactic IC emission is not well-understood, and it
is unlikely that one can make a definitive statement as to whether or
not IC is the emission mechanism without extensive modeling and further
analysis.  In a very recent paper, \citeasnoun{moskalenko} shows that
IC in the framework of a hard electron spectrum is a viable candidate
for explaining the general Galactic plane excess above 1 GeV.  
This paper additionally indicates that, at least for
a longitudinally averaged profile,
a model with a halo a few kpc in height reasonably accounts for the 
latitude profile up to high latitudes, from low to high energies
(100 MeV - 1 GeV).
The obvious next step (which we shall explore for a future paper)
is to perform TIPSH analysis using a 2D version of this model (derived
from a cosmic-ray propagation code) as the null hypothesis.  Also,
we note that IC profiles typically have a strong component in the
Galactic plane (mostly in the GC region) in addition to high-latitude 
emission.  If the standard model underestimates IC, then model fits
would result in an {\em overestimate} of the contribution from
cosmic-ray/matter interations.  This may explain our results at
lower energies, which show a systematically negative residual in the
Galactic plane away from GC, but still strong evidence for 
large-scale high-latitude emission surrounding the GC region.

Some models of particle dark matter, both baryonic 
(\citeasnoun{depaolis})
and non-baryonic 
(\citeasnoun{jungman}), predict $\gamma$-ray emission as
an annihilation or interaction product.  Annihilations of
pairs of weakly interacting massive particles (WIMPs) are
expected to generate $\gamma$-rays (among other things) as final
products (\citeasnoun{jungman}).  
The ``smoking gun'' in this case would be the detection
of a high-energy line at the WIMP mass, due to the processes
$\chi\chi \rightarrow \gamma\gamma$ and $\chi\chi \rightarrow Z\gamma$
(\citeasnoun{ullio}).
We see no evidence for this below 10 GeV.  A continuum spectrum is
also expected as secondary products from the reaction
$\chi\chi \rightarrow q\bar{q}$, however the signature of this
continuum would be more difficult to detect.

The key question here is whether or not WIMP annihilation is at
all a realistic candidate.  The answer is a qualified ``yes'', with
our result providing some constraints.  If we assume a generic
Galactic distribution of WIMPs which is smooth (no clumping, see
e.g. \citeasnoun{turner} and \citeasnoun{kamion}), we can arrive
at a rough estimate of $\langle \sigma v \rangle$, the average of
the product of the WIMP annihilation cross-section 
and relative velocities.
For our result this turns out to be something like
$O(10^{-25})-O(10^{-24})$.  If we ignore exotic phenomena, and assume
for simplicity that the annihilation occurs only via the s-wave channel,
the relic WIMP density is given by (\citeasnoun{jungman})
\begin{equation}
\Omega_\chi h^2 \simeq \frac{3\times 10^{-27} {\mathrm cm^3 s^{-1}}}
	{\langle \sigma_A v \rangle},
\end{equation}
where the Hubble constant is given as $100 h$ km s$^{-1}$ Mpc$^{-1}$
and $\sigma_A$ is the total WIMP annihilation cross-section.
For the ``standard'' cosmological model with $\Omega = 1, h \simeq 1$,
the annihilation cross-section above is far too large for WIMPs
to account for an interesting quantity of dark matter (see also
\citeasnoun{cline}).  One way out
of this is to have clumpy dark matter, as discussed in 
\citeasnoun{wasserman}, \citeasnoun{gurevich},
and in more recently in detail for large classes of supersymmetric (SUSY)
WIMP models in \citeasnoun{bergstrom1} and \citeasnoun{bergstrom2}.  
The reason
for this is that the $\gamma$-ray intensity goes as the {\em square}
of the WIMP density, and so modest local density enhancements can
lead to large increases in $\gamma$-ray emissivity.  In particular,
\citeasnoun{bergstrom2} notes that it is the combination of the
$\gamma$-ray and cosmic-ray antiproton intensities which place the
best constraints on SUSY and clumping parameters.  Another way to
make WIMPs more viable is to modify the cosmological model.  In fact,
recent results from high-redshift supernova surveys
(\citeasnoun{filippenko}) indicate
$h \simeq 0.65$ and $\Omega_{matter} \simeq 0.3$, which brings
the implied annihilation cross-section at least closer to what
we calculate.  In fact, \citeasnoun{gondolo} has shown
that one can explain the observed halo quite nicely following a similar
approach.  So this remains an open question.

Finally, we come to the baryonic dark matter hypothesis.
A number of authors have proposed that Galactic dark matter may
be in the form of cold, dark molecular clouds (see e.g.
\citeasnoun{pfenniger}, \citeasnoun{gerhard}).
\citeasnoun{walkerwardle} have postulated that extreme scattering
events in the radio may be attributed to photo-ionized electrons
in such clounds as they pass through the line of sight.  \citeasnoun{walker}
notes that collisions amongst such clouds naturally leads to a cored
halo distribution, and may form the physical basis for the Tully-Fisher
relation.
De Paolis \etal\ 1995 also suggest that a significant portion of dark matter
may come in the form of clusters of MACHOs and/or cold H$_2$ molecular
clouds.  If in the form of cold H$_2$ clouds, then we expect some
$\gamma$-ray emission resulting from cosmic ray protons and the
reactions $pp \rightarrow \pi^0 \rightarrow \gamma\gamma$.
De Paolis \etal\ make an estimate of the distribution of high latitude
cosmic ray protons, and predict the $\gamma$-ray flux above 1 GeV
at the Galactic poles to be 
$\Phi_\gamma(90^\circ) \simeq \epsilon 1.7\times 10^{-6}$
cm$^{-2}$ s$^{-1}$ sr$^{-1}$, where $\epsilon$ is some efficiency factor
included to account for uncertainties in the CR proton distribution.

Though exposure artifacts at the poles prevent us from making a direct
comparison, this estimate is clearly in the neighborhood of what we
have observed.  One question which arises in this scenario is whether
or not one expects to see many $\gamma$-rays: if the CR protons have
largely outward momenta, so too will the $\gamma$-rays, resulting in a
small $\epsilon$.  In order to get the $\gamma$-ray flux we observe,
some sort of trapping of CR protons in the Galactic halo must occur.
\citeasnoun{simpson} has recently reported that, based on measurements
of isotopic abundances in cosmic-rays (e.g., $^{26}$Al/$^{27}$Al),
that cosmic-ray lifetimes are perhaps a factor of four larger than
previously thought, and that the cosmic-rays are traversing an average ISM
density smaller than that observed for the Galaxy.  
\cite{strong4} also report, based on $^{10}$Be/$^9$Be measurements
and comparison with cosmic-ray propagation models, that the height
of the halo propagation regions is $>4$ kpc.
One interpretation
of this is that cosmic-rays are trapped in the low-density halo, making
the \citeasnoun{depaolis} scenario somewhat 
more appealing.
There are clearly large uncertainties associated
with this hypothesis, but it seems an intriguing avenue for further study.

\section{Future Directions}
The establishment of the existence and (to some extent) the spatial
distribution of the halo indicates several lines of future research.
Clearly one step to be taken in the near future is to do the TIPSH
analysis on data taken over smaller energy bins, to try and establish
(at least qualitatively) the energy dependence of the halo morphology.
We also note that the Galactic diffuse model used here is only one
choice amongst many.  Future analyses will use different and more
sophisticated hypotheses, as well as include the effects of point
sources in the null hypothesis.
Another direction is in the application on parametric analyses to the
same data to fill in that information which is difficult to obtain
non-parametrically.  This includes quantitative estimates of fluxes
and errors, as well as parameterizations of the halo morphology as
dictated by various physical models.  Hypotheses as to the origin
of the halo can potentially be rejected on the basis of spatial/spectral
behavior.  Also, if the halo truly is Galactic, estimates of the
extragalactic $\gamma$-ray background may be impacted.  Galactic halo 
models which give rise to $\gamma$-ray emission not only include a
bulge-like component as we've seen here, but generally larger
scale emission as well.  While such emission may not be precisely
isotropic, it is typically dim (compared to the bulge component) and
spatially more slowly varying; as such it may be statistically
indistinguishable from isotropic emission in the EGRET data.  
In our very simple parametric
fit above, the estimate of the isotropic emission changed by
$\sim$25\% depending on whether or not a halo component was included,
so this possibility is in no way negligible.

We saw above that the key limitation to understanding (at least
non-parametrically) the spatial distribution of the halo is the
highly non-uniform exposure of EGRET observations.  It is unlikely
this will improve much, given that EGRET is basically out of spark
chamber gas.  In principle we could include more observations from
later CGRO cycles, but these contain rather little additional information,
and generally don't serve to even out the exposure.  The upcoming
GLAST mission may provide better data for these types of studies.
GLAST will be more sensitive over a larger energy band, and further, if
operated in a scanning (as opposed to pointed) mode, will yield an all-sky
exposure considerably more uniform than EGRET's, thus removing a
key systematic.  Detailed understanding of what GLAST may be
able to do in this context remains the subject of future studies.

\section{Conclusions}
We have presented strong statistical evidence for a large scale anisotropic
excess in high energy $\gamma$-rays.  Examination of our maps indicates
that this excess may originate in the Galactic halo, though our results
in no way rule out a local origin.  The spectrum appears to be broadband;
detailed investigation of the spectral properties will be the subject
of future work.  The origin of the halo is unclear.  
Emission from inverse Compton in a large halo \cite{moskalenko} 
appears to be a good candidate but it remains to be seen whether it can 
account for the entire observations in detail.
A definitive answer
on this topic will perhaps require a ``smoking gun'' in this or 
another energy
band, or may have to wait for the GLAST mission.  Further, as noted
by previous authors (\citeasnoun{smialkowski}; 
\citeasnoun{chen}; 
\citeasnoun{strong1}),
the existence of such an extended excess may impact estimates of
the extragalactic $\gamma$-ray background.

\section{Reproducible Research, Color Figures}
Data and software to reproduce the results of this paper can be found
at\\
\url{http://tigre.ucr.edu/halo/repro.html}.

Color versions (in JPEG format) of the figures are available at\\
\url{http://tigre.ucr.edu/halo/paper.html}.

\section*{Acknowledgements}
DDD and DHH gratefully acknowledge the support of the 
Max-Planck-Gesellschaft.  DDD and JS were supported for this work 
by the NASA CGRO Guest Investigator program.

Much of the data used in the analysis for this paper was provided
by the Compton Observatory Science Support Center, with the help
of D. Macomb.  We also thank P. Sreekumar for his help in obtaining
data and in understanding the results, as well as D. Cline, F. De Paolis,
and P. Gondolo for helpful discussions on the possible origins of
the halo.  We thank E. Bloom for input on the proton initiated background
in EGRET.

DDD dedicates this paper to his grandfather, A. James Ebel, a lifelong
supporter of science.

\newpage
\figcaption[haarfig.ps]{
Two-dimensional Haar wavelets, encoding image variations in the
a) horizontal, b) vertical, and c) diagonal directions.
\label{fig:haar}
}

\figcaption[diffmod.ps]{
Galactic diffuse model used in the null hypothesis (isotropic component
is not included), truncated at the same flux as
Figure \protect{\ref{fig:halo}}.  Contour values are flux in units of
ph cm$^{-2}$ s$^{-1}$ sr$^{-1}$.
\label{fig:diffmod}
}

\figcaption[halo1_GeV_0_p.ps,halo1_GeV_0_m.ps]{
Contours are in units of $10^{-5}$ ph cm$^{-2}$ s$^{-1}$ sr$^{-1}$.
a) Positive TIPSH residual for $E>1$ GeV, unscaled Galactic diffuse model.
Fluxes have been truncated at $10^{-5}$ ph cm$^{-2}$ s$^{-1}$ sr$^{-1}$.
b) Negative TIPSH residual as for (a).  Flux truncated at
$-2\times 10^{-6}$ ph cm$^{-2}$ s$^{-1}$ sr$^{-1}$.
\label{fig:halo}
}

\figcaption[halo1_GeV_40_p.ps,halo1_GeV_40_m.ps]{
Contours are in units of $10^{-5}$ ph cm$^{-2}$ s$^{-1}$ sr$^{-1}$.
a) Positive TIPSH residual for $E>1$ GeV, Galactic diffuse model
scaled by 1.4.
Fluxes have been truncated at $10^{-5}$ ph cm$^{-2}$ s$^{-1}$ sr$^{-1}$.
b) Negative TIPSH residual as for (a).  Flux truncated at
$-2\times 10^{-6}$ ph cm$^{-2}$ s$^{-1}$ sr$^{-1}$.
\label{fig:40}
}

\figcaption[halo1_GeV_20_p.ps,halo1_GeV_20_m.ps]{
Contours are in units of $10^{-5}$ ph cm$^{-2}$ s$^{-1}$ sr$^{-1}$.
a) Positive TIPSH residual for $E>1$ GeV, Galactic diffuse model
scaled by 1.2.
Fluxes have been truncated at $10^{-5}$ ph cm$^{-2}$ s$^{-1}$ sr$^{-1}$.
b) Negative TIPSH residual as for (a).  Flux truncated at
$-2\times 10^{-6}$ ph cm$^{-2}$ s$^{-1}$ sr$^{-1}$.
\label{fig:20}
}

\figcaption[ic1_GeV.ps,ric1_GeV.ps]{
Contours evenly spaced between 0 and 
$10^{-5}$ ph cm$^{-2}$ s$^{-1}$ sr$^{-1}$.
a) Model of inverse Compton emission above 1 GeV.
b) TIPSH recovered flux distribution for simulated data including
(a).
\label{fig:ic}
}

\figcaption[exp1_GeV.ps]{
EGRET exposure for CGRO Phases 1-4.
\label{fig:exposure}
}

\figcaption[halo_exp1_GeV_20_p.ps]{
Truncated denoised residual with Phase 1-4 exposure overplotted as
dotted contours.
\label{fig:haloexp}
}

\figcaption[sim1_GeV_0.ps]{
TIPSH residual for simulated data based on the predicted diffuse emission.
Units are counts.
\label{fig:sim0}
}

\figcaption[flat1_GeV_100.ps]{
Fractional deviation in TIPSH estimated flux for simulated data
consisting of an isotropic excess at the level of the measured
isotropic background.  The exposure factor is overplotted as
contours.
\label{fig:flat}
}

\figcaption[sim1_GeV_40.ps]{
Contours are in units of $10^{-5}$ ph cm$^{-2}$ s$^{-1}$ sr$^{-1}$.
a) 40\% scaled standard Galactic diffuse model.
TIPSH residual for simulated data based on the predicted emission
with an excess of 40\% Galactic diffuse, using the standard prediction
as the null hypothesis.
\label{fig:sim40}
}

\figcaption[icrat1_GeV.ps]{
Ratio of the inverse Compton model in Figure \ref{fig:ic}a
to the recovered flux distribution in  Figure \ref{fig:ic}b.
\label{fig:icrat}
}

\figcaption[halo100_300MeV_0_p.ps,halo100_300MeV_0_m.ps]{
Contours are in units of $10^{-5}$ ph cm$^{-2}$ s$^{-1}$ sr$^{-1}$.
a) Positive TIPSH residual for $100<E<300$ MeV, 
unscaled Galactic diffuse model.
Fluxes have been truncated at 
$2\times 10^{-5}$ ph cm$^{-2}$ s$^{-1}$ sr$^{-1}$.
b) Negative TIPSH residual as for (a).  Flux truncated at
$-10^{-5}$ ph cm$^{-2}$ s$^{-1}$ sr$^{-1}$.
\label{fig:100}
}

\figcaption[halo300_1000MeV_0_p.ps,halo300_1000MeV_0_m.ps]{
Contours are in units of $10^{-5}$ ph cm$^{-2}$ s$^{-1}$ sr$^{-1}$.
a) Positive TIPSH residual for $300<E<1000$ MeV, 
unscaled Galactic diffuse model.
Fluxes have been truncated at 
$1.5\times 10^{-5}$ ph cm$^{-2}$ s$^{-1}$ sr$^{-1}$.
b) Negative TIPSH residual as for (a).  Flux truncated at
$-5\times 10^{-6}$ ph cm$^{-2}$ s$^{-1}$ sr$^{-1}$.
\label{fig:300}
}

\figcaption[albedo1_GeV_20_p.ps]{
The dark curves bound regions where Earth albedo contamination is
expected to be greatest.  Note no particular excess in these regions,
nor correlation with the spatial distribution of the halo.
\label{fig:albedo}
}

\newpage
\mbox{ }
\newpage
\begin{table}
\begin{tabular}{||c||c|c|c||c|c|c||} \hline
Wavelet scale & 
\multicolumn{3}{c||}{No. wavelets det. (data)} & 
\multicolumn{3}{c||}{No. wavelets det. (null)}\\ \hline
(pixels) & Hor & Vert & Diag & Hor & Vert & Diag\\
\hline\hline
2 & 336 & 325 & 190 & 0 & 0 & 0\\ \hline
4 & 846 & 965 & 537 & 0 & 0 & 0\\ \hline
8 & 2307 & 2831 & 1783 & 0 & 0 & 0\\ \hline
16 & 7864 & 8387 & 5767 & 
	0 & 0 & 0\\ \hline
32 & 25166 & 26948 & 20552 & 
	0 & 0 & 0\\ \hline
64 & $1.0\times 10^5$ & $1.1\times 10^5$ & 78538 & 
	8 & 0 & 0\\ \hline
128 & $3.3\times 10^5$ & $3.9\times 10^5$ & $3.1\times 10^5$ & 
	1 & 0 & 0 \\ \hline
256 & $7.8\times 10^5$ & $9.2\times 10^5$ & $7.4\times 10^5$ & 
	0 & 0 & 0 \\ \hline
512 & $7.9\times 10^5$ & $8.0\times 10^5$ & $6.1\times 10^5$ & 
	0 & 0 & 0 \\ \hline
\end{tabular}
\caption{Number of wavelets detected (i.e., those that passed
threshhold) for actual data (in the case given in Figure \ref{fig:20})
and for simulated data generated by the null hypothesis, as a function
of the wavelet direction (horizontal, vertical, or diagonal).  The level-wise
false detection rate (FDR) was set at $10^{-4}$.  Note that the number
of detections for the data is much higher than that for the null,
which is much more in line with what we expect for the chosen FDR.  
The detection rates above are averaged over
the image shifts.  This adds a slight subtlety, since the tests
at different shifts are {\em not} independent, so a wavelet that
is detected in one shift may very well also be detected in nearby
shifts.}
\label{tab:rates}
\end{table}


\begin{thebibliography}{}

\bibitem[Bergstr\"om, Edsj\"o, and Ullio 1998]{bergstrom1}
Bergstr\"om, L., Edsj\"o, J., and Ullio P. 1998,
\url{http://xxx.lanl.gov/abs/astro-ph/9804050}.

\bibitem[Bergstr\"om \etal. 1998]{bergstrom2}
Bergstr\"om, L., \etal. 1998, \url{http://xxx.lanl.gov/abs/astro-ph/9806072}.

\bibitem[Bertsch \etal. 1993]{bertsch}
Bertsch, D. L., \etal. 1993, ApJ, 416, 587.
\bibcode{1993ApJ...416..587B}

\bibitem[Chen, Dwyer, and Kaaret 1996]{chen}
Chen, A., Dwyer, J., and Kaaret, P., 1996, ApJ, 463, 169.
\bibcode{1996ApJ...463..169C}

\bibitem[Cline 1998]{cline}
Cline, D. B. 1998, ApJ, 501L, 1.
\bibcode{1998ApJ...501L...1C}

\bibitem[De Paolis \etal. 1995]{depaolis}
De Paolis, F. \etal., 1995, A\&A, 295, 567.
\bibcode{1995A&A...295..567D}

\bibitem[Dixon \etal. 1997]{dixon1}
Dixon, D. D., \etal. 1997, Proc. 4th Compton Symposium, AIP, 410, 1198.
\bibcode{1997AIPC..410.1601D}

\bibitem[Dixon \etal. 1998]{dixon2}
Dixon, D. D., \etal. 1998, Proc. Third. Int. Symp.
on Sources and Detection of Dark Matter in the Universe, ed. D. Cline
(Amsterdam: Elsevier), in press.

\bibitem[Donoho 1995]{donoho}
Donoho, D. L., 1995, IEEE Trans. Information Theory, 41, 613.
\bibcode{1995ITIT...41..613D}

\bibitem[Donoho and Coifman 1995]{donocoif}
Donoho, D.L and Coifman, R.R., 1995, in {\it Wavelets and Statistics,} 
ed. Antoniadis, A. and Oppenheim, G. (Springer-Verlag).


\bibitem[Filippenko and Riess 1998]{filippenko}
Filippenko, A. V. and Riess, A. G. 1998, Proc. Third. Int. Symp.
on Sources and Detection of Dark Matter in the Universe, ed. D. Cline
(Amsterdam: Elsevier), in press.

\bibitem[Gerhard and Silk 1996]{gerhard}
Gerhard, O. and Silk, J. 1996, ApJ, 472, 34.
\bibcode{1996ApJ...472...34G}

\bibitem[Gondolo 1998]{gondolo}
Gondolo, P. 1998, \url{http://xxx.lanl.gov/abs/astro-ph/9807347}.

\bibitem[Gurevich and Zybin 1997]{gurevich}
Gurevich, A. V. and Zybin, K. P., 1997, Phys. Lett. A, 225, 217.
\bibcode{1997PhLA..225..217G}

\bibitem[Hartmann 1995]{hartmann}
Hartmann, D. H., 1995, ApJ, 447, 646.
\bibcode{1995ApJ...447..646H}

\bibitem[Hunter \etal. 1997]{hunter}
Hunter, S. D., \etal. 1997, ApJ, 481, 205.
\bibcode{1997ApJ...481..205H}

\bibitem[Jungman, Kamionkowski, and Griest 1996]{jungman}
Jungman, G., Kamionkowski, M., and Griest, K., 1996, Physics Reports, 267, 195.
\bibcode{1996PhR...267..195J}

\bibitem[Kamionkowski and Kinkhabwala 1997]{kamion}
Kamionkowski, M. and Kinkhabwala, A., 1998, Phys. Rev. D, 57, 3256.
\bibcode{1998PhRvD..57.3256K}

\bibitem[Kolaczyk 1996]{kola1}
Kolaczyk, E. D., 1996, Technical Report, Department of Statistics,
University of Chicago.

\bibitem[Kolaczyk 1997]{kola2}
Kolaczyk, E. D., 1997, ApJ, 483, 340.
\bibcode{1997ApJ...483..340K}

\bibitem[Kolaczyk and Dixon 1998]{kola3}
Kolaczyk, E. D. and Dixon, D. D., 1998, in preparation.

\bibitem[Lamb and Macomb 1997]{lamb}
Lamb, R. C. and Macomb, D. J. 1997, ApJ, 488, 872.
\bibcode{1997ApJ...488..872L}

\bibitem[Longair 1992]{longair}
Longair, M. S., 1992, {\em High energy astrophysics}
(New York: Cambridge University Press).

\bibitem[Mattox \etal. 1996]{mattox}
Mattox, J. R., \etal. 1996, ApJ, 461, 396.
\bibcode{1996ApJ...461..396M}

\bibitem[Moskalenko and Strong 1998]{moskalenko}
Moskalenko, I. V. and Strong, A. W. 1998,
Proceedings of the 16th European Cosmic Ray
Symposium (July 1998, Alcala), in press.
\url{http://xxx.lanl.gov/abs/astro-ph/9807288}

\bibitem[Page and Hawking 1976]{page}
Page, D. N. and Hawking, S. W. 1976, ApJ, 206, 1.
\bibcode{1976ApJ...206....1P}

\bibitem[Pfenniger, Combes, and Martinet 1994]{pfenniger}
Pfenniger, D., Combes, F., and Martinet, L. 1994, A\&A, 285, 79.
\bibcode{1994A&A...285...79P}

\bibitem[Pohl 1998]{pohl}
Pohl, M. 1998, Proceedings of the 16th European Cosmic Ray
Symposium (July 1998, Alcala), in press. 
\url{http://xxx.lanl.gov/abs/astro-ph/9807267}

\bibitem[Posten 1989]{posten}
Posten, H.O., 1989, The American Statistician, 43, 261.

\bibitem[Simpson 1998]{simpson}
Simpson, J. A. and Connell, J. J. 1998, ApJ, 497L, 85.
\bibcode{1998ApJ...497L..85S}

\bibitem[Smialkowski, Wolfendale, and Zhang 1997]{smialkowski}
Smialkowski, A., Wolfendale, A. W., Zhang, L., 1997, 
Astroparticle Phys, 7, 21.
\bibcode{1997APh.....7...21S}

\bibitem[Sreekumar \etal. 1998]{sreekumar}
Sreekumar, P. \etal., 1998, ApJ, 494, 523.
\bibcode{1998ApJ...494..523S}

\bibitem[Strong \etal. 1983]{strong1}
Strong, A. W. \etal., 1983, 18th International Cosmic Ray Conference,
Bangalore, India, ed. N. Durgaprasad \etal. , 9, 90.
\bibcode{1983icrc....9...90S}

\bibitem[Strong 1984]{strong2}
Strong, A. W., 1984, Adv. Space Res., 3, 87
\bibcode{1984AdSpR...3...87S}

\bibitem[Strong and Youssefi 1995]{stryou}
Strong, A. W. and Youssefi, G., 1995, Proc. 24th ICRC, 3, 48.
\bibcode{1995icrc....3...48S}


\bibitem[Strong and Moskalenko 1998]{strong4}
Strong, A. W. and Moskalenko, I. V. 1998, ApJ, 509, in press.
\url{http://xxx.lanl.gov/abs/astro-ph/9807150}

\bibitem[Turner 1986]{turner}
Turner, M. S., 1986, Phys. Rev. D, 34, 1921.
\bibcode{1986PhRvD..34.1921T}

\bibitem[Ullio and Bergstrom 1998]{ullio}
Ullio, P. and Bergstrom, L., 1998, Phys. Rev. D, 57, 1962.
\bibcode{1998PhRvD..57.1962U}

\bibitem[Walker and Wardle 1998]{walkerwardle}
Walker, M. and Wardle, M. 1998, ApJ, 498L, 125.
\bibcode{1998ApJ...498L.125W}

\bibitem[Walker 1998]{walker}
Walker, M. A. 1998, MNRAS, submitted.
\url{http://xxx.lanl.gov/abs/astro-ph/9807236}

\bibitem[Wasserman 1996]{wasserman}
Wasserman, I., 1996, \url{http://xxx.lanl.gov/abs/astro-ph/9608012}.

\bibitem[Wright 1996]{wright}
Wright, E. L. 1996, ApJ, 459, 487.
\bibcode{1996ApJ...459..487W}

\bibitem[Willis 1996]{willis}
Willis, T. D., 1996, Ph.D. Thesis, Dept. of Physics, Stanford University.

\end{thebibliography}
\end{document}